\documentclass[aps,twocolumn,showpacs,nofootinbib]{revtex4-1}
\usepackage{graphicx}   
\usepackage{latexsym}   
\usepackage{amsmath}

\usepackage[normalem]{ulem}  
\usepackage[dvips]{color} 

\renewcommand\sout{\bgroup \color{red} \ULdepth=-.5ex \ULset}

\begin{document}
\title{Deep sub-threshold $\Xi$ and $\Lambda$ production in nuclear collisions with the UrQMD transport model}

\author{G. Graef$^1$, J. Steinheimer$^1$, Feng Li$^{2,3}$, and Marcus Bleicher$^{1,4}$}

\affiliation{$^1$Frankfurt Institute for Advanced Studies, Ruth-Moufang-Str. 1, 60438 Frankfurt am Main, Germany}
\affiliation{$^2$Cyclotron Institute, Texas $A\&M$ University, College Station, Texas 77843, USA}
\affiliation{$^3$Department of Physics and Astronomy, Texas $A\&M$ University, College Station, Texas 77843, USA}
\affiliation{$^4$Institut f\"ur Theoretische Physik, Goethe-Universit\"at, Max-von-Laue-Str. 1, 60438 Frankfurt am Main, Germany}

\begin{abstract}
We present results on deep sub threshold hyperon production in nuclear collisions, with the UrQMD transport model. Introducing anti-kaon+baryon and hyperon+hyperon strangeness exchange reactions we obtain a good description of experimental data on single strange hadron production in Ar+KCl reactions at $E_{lab}=1.76$ A GeV. We find that the hyperon strangeness exchange is the dominant process contributing to the $\Xi^-$ yield, however remains short of explaining the $\Xi^-/\Lambda$ ratio measured with the HADES experiment. We also discuss possible reasons for the discrepancy with previous studies and the experimental results, finding that many details of the transport simulation may have significant effects on the final $\Xi^-$ yield.
\end{abstract}

\maketitle
\section{Introduction}
The study of strange hadrons (hyperons) has always been a focus in dedicated heavy ion experiments. Since strangeness has to be produced as $s+\overline{s}$ pair, its production mechanism may give us insight in the properties of the matter produced at relativistic nuclear collisions \cite{Rafelski:1982pu,Koch:1986ud}. In p+p collisions there is a strict energy threshold for the production of hyperons, given by the sum of the hyperon and kaon masses. In collisions of nuclei, with a significant number of secondary reactions, this threshold is essentially lifted \cite{Randrup:1980qd} and the rate of strangess production can help us understand the collision dynamics. In several previous works it was indicated that the production rates and properties of kaons, in low energy nuclear collisions, are a promising probe to extract the medium interactions of kaons \cite{Aichelin:1986ss,Shor:1989nz,Hartnack:1993bq,Fang:1994cm,Li:1994cu,Li:1994vy,Mosel:1992rb,Miskowiec:1994vj,Cassing:1996xx,Bratkovskaya:1997pj,Hartnack:2001zs,Hartnack:2005tr,Hartnack:2011cn}. At high beam energies $\sqrt{s_{NN}}> 5$ GeV the production rate of strange particles, in central heavy ion collisions, is well estimated by a grand canonical fit to the particle yields (see e.g. \cite{Becattini:2000jw,Andronic:2005yp}), indicating that strangeness may approach chemical equilibrium. At lower energies canonical effects and/or finite size effects have to be taken into account to explain the yield of strange hadrons. An important question is whether this equilibration of strangeness is directly related to the onset of deconfinement, or if hadronic interactions are sufficient to equilibrate the chemical composition of the produced fireball. Microscopic transport models generally underestimate the yield of strange (and especially multi-strange) hadrons, at higher collision energies $\sqrt{s_{NN}}\gtrsim 5$ GeV, but it is unclear if this is merely due to missing hadronic strangeness production reactions. Important for the understanding of the production of strange hadrons through hadronic channels, are nuclear reactions where strange hadrons are produced below the threshold energy of the corresponding elementary p+p collisions. The threshold center-of-mass energies in p+p collisions are $m_p+m_K+m_{\Lambda}=2548$ GeV for $\Lambda$ production and $m_p+2m_K+m_{\Xi}=3240$ GeV for $\Xi$ production. Here, the production rate depends strongly on multiple secondary interactions and is therefore escpecially sensitive to, hadronic production channels. The HADES \cite{Agakishiev:2009rr,Agakishiev:2010rs,Agakishiev:2009ar,Tlusty:2009dk} and FOPI \cite{Ritman:1995tn,Herrmann:1996zg,Crochet:2000fz,Best:1997ua} experiments at the SIS18 accelerator have recently measured strange particle yields from collisions of nuclei at beam energies of $E_{lab}= 1.76$ and $1.23$ A GeV, supplementing earlier KAOS measurements \cite{Forster:2007qk}.\\
The purpose of this paper is to show how the hadronic transport model UrQMD can be extended to include most relevant strange hadron production mechanisms used to calculate strange particle ratios at beam energies where hyperons are produced below their p+p threshold. The results from the model will then be compared to experimental data and a thermal model fit to the data. We will also make predictions for strange particle yields expected for Au+Au collisions at a beam energy of $E_{lab}=1.23$ A GeV.

\section{Strangeness exchange reactions in UrQMD}
To estimate the production of strange particles through hadronic channels, we will employ the hadronic Ultra-relativistic Quantum Molecular Dynamics model (UrQMD) \cite{Bass:1998ca,Bleicher:1999xi}.
The model is based on an effective microscopic solution of the relativistic Boltzmann equation \cite{DeGroot:1980dk}:
\begin{equation}
\label{boltzmann}
p^\mu \cdot \partial_\mu f_i(x^\nu, p^\nu) = \mathcal{C}_i \quad .
\end{equation}

This equation describes the time evolution of the distribution functions for particle species $i$ and includes the full collision term on the right hand side.
It includes 39 different hadronic species (and their anti-particles) which scatter according to their geometrical cross section. The allowed processes include elastic scatterings and $2\rightarrow n$ processes via resonance creation (and decays) as well as string excitations for large center-of-mass energies ($\sqrt{s}\gtrsim 3$ GeV) .\\
Previous versions of the model did not include strangeness exchange processes like the exchange of the strange quark from a strange meson to a nucleus, i.e. $\overline{K}+N \leftrightarrow \pi +Y$ (where $Y$ is a strange Baryon) which have been found to be important in other transport model calculations \cite{Ko:1984nv,Cassing:1996xx,Hartnack:2001zs,Chen:2003nm}. These reactions are measured by experiment and we include them in the UrQMD model, using cross sections parametrized by a fit to available experimental data \cite{Flaminio:1983fw}: 
\begin{eqnarray}
	\sigma_{K^- + p \rightarrow \pi^- + \Sigma^+} &=&  \frac{0.0788265}{(\sqrt{s}-1.38841 \rm{GeV})^2}, \nonumber \\
	\sigma_{K^- + p \rightarrow \pi^+ + \Sigma^-} &=&  \frac{0.0196741}{(\sqrt{s}-1.42318 \rm{GeV})^2}, \nonumber \\	
	\sigma_{K^- + p \rightarrow \pi^0 + \Sigma^0} &=&  \frac{0.55 \cdot 0.0508208}{(\sqrt{s}-1.38837 \rm{GeV})^2},\nonumber \\
  \sigma_{K^- + p \rightarrow \pi^0 + \Lambda}  &=&  \frac{0.45 \cdot 0.0508208}{(\sqrt{s}-1.38837 \rm{GeV})^2}.
	\end{eqnarray}

\begin{figure}[t]	
\center \includegraphics[width=0.5\textwidth]{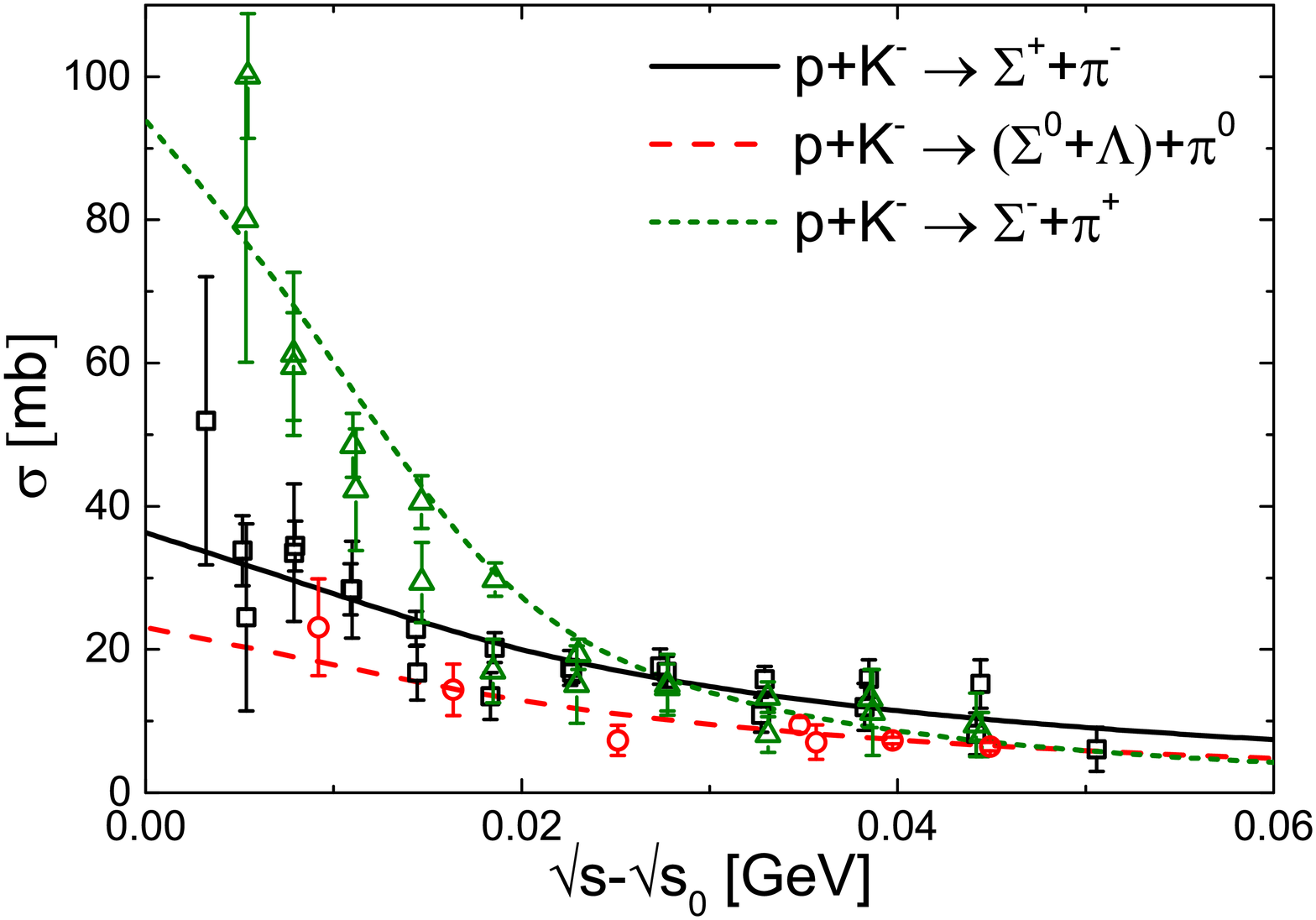}	
\caption{(Color Online) Kaon-nucleon strangeness exchange cross sections implemented in the UrQMD transport model (lines) together with experimental data (symbols) from \cite{Flaminio:1983fw}. }
\label{f1}
\end{figure}

The parametrized cross sections together with the data are shown in figure \ref{f1}. All other possible iso-spin channels and the necessary back reaction follow from isospin symmetry and detailed balance relations which are implemented in the most recent version of the UrQMD transport model (UrQMD v3.4 \footnote{www.urqmd.org}).

In nuclear collisions, at low beam energies, Baryon-Baryon interactions play an important role in the description of the collision dynamics. We therefore expect that another important strangeness exchange reaction, the two hyperon exchange $Y+Y\rightarrow \Xi + N$, will be important for the production of doubly strange hyperons \cite{Li:2012bga}. As there exist no direct measurements of these reaction cross sections, we have to employ an effective model to estimate the associated cross sections. 
In the present study we will use the cross sections provided by Feng Li et. al. \cite{Li:2012bga} and the back reaction is calculated according to detailed balance.
Feng Li et. al. employed a gauged flavor SU(3)-invariant hadronic Lagrangian \cite{Li:2002yd}, and calculated the cross sections for
the strangeness-exchange reactions $Y+Y\rightarrow N+\Xi$ in the Born approximation.\\

\begin{figure}[t]	
\center \includegraphics[width=0.5\textwidth]{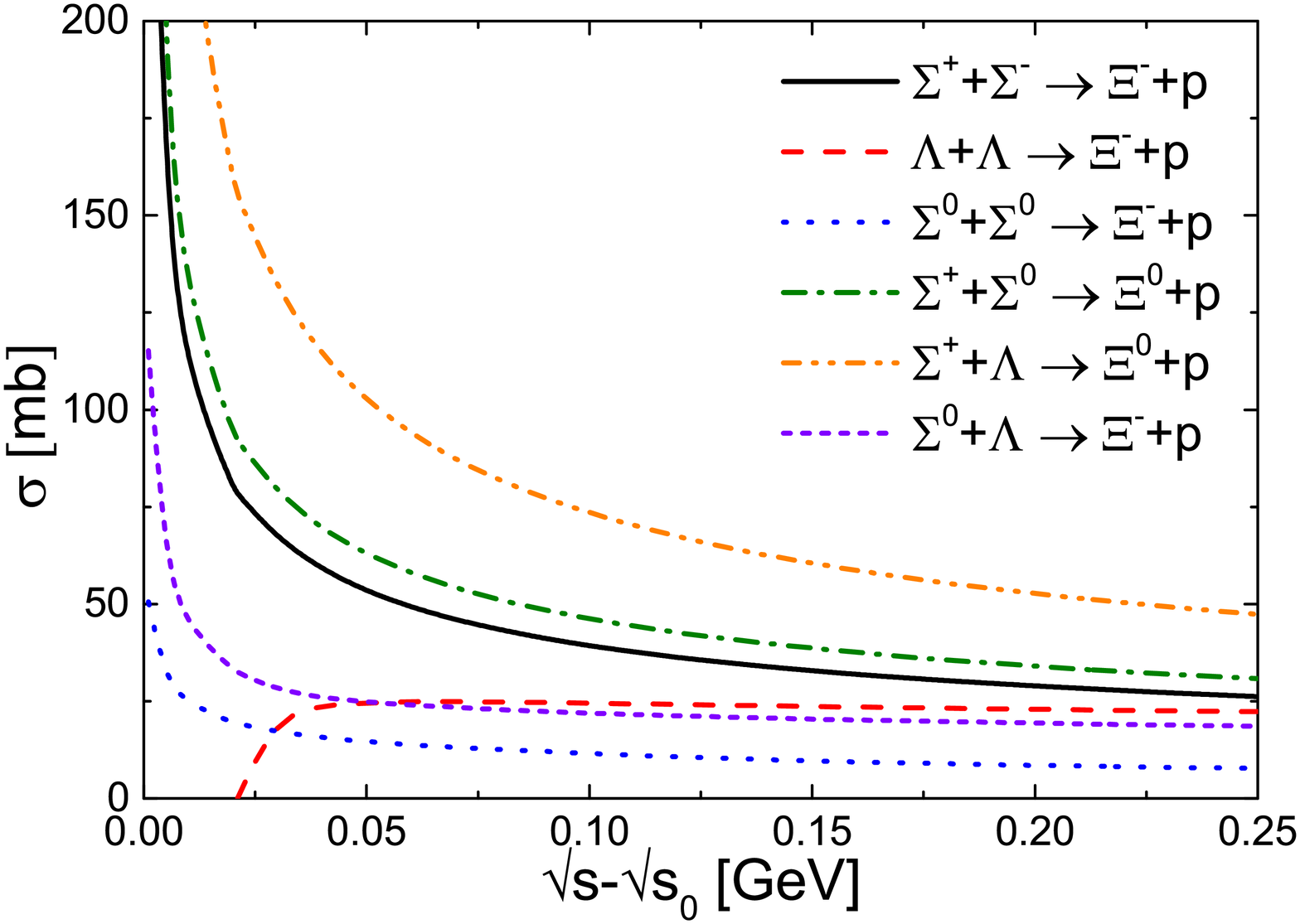}	
\caption{(Color Online) Parametrized iso-spin dependent cross sections, for the hyperon+hyperon strangeness exchange, used in the UrQMD transport model simulation \cite{Li:2012bga}. }
\label{f2}
\end{figure}

Since UrQMD is a microscopic transport model with explicit local conservation of energy, momentum and charge, we need to implement iso-spin dependent crossections of the hyperon+hyperon strangeness exchange reactions. These can also be inferred from the model used in \cite{Li:2012bga}.
The specific isospin dependent cross sections for  $Y+Y\rightarrow \Xi + N$  are then given by
\begin{eqnarray}\label{crs}
\sigma_{\Lambda Y\to N\Xi}(I_Y,I_N,I_\Xi,s) = \frac{1}{64\pi sp_i^2}\frac 1 {(2s_1+1)(2s_2+1)}\nonumber\\
\times\sum_{s_{1}s_{2}s_{1}^{\prime }s_{2}^{\prime }}\int dt\overline{\left\vert \mathcal{M}_{s_{1}s_{2}s_{1}^{\prime}s_{2}^{\prime}}(I_Y,I_N,I_\Xi)\right\vert^{2}}\nonumber\\
\sigma_{\Sigma\Sigma\to N\Xi}(I_1,I_2,I_N,I_\Xi,s)=\frac{1}{64\pi sp_i^2}\frac 1 {(2s_1+1)(2s_2+1)}\nonumber\\
\times\frac 1 2\sum_{s_{1}s_{2}s_{1}^{\prime }s_{2}^{\prime }}\int dt \left[ \overline{\left\vert \mathcal{M}_{s_{1}s_{2}s_{1}^{\prime}s_{2}^{\prime}}(I_1,I_2,I_N,I_\Xi)\right\vert^{2}} \right. \nonumber \\
+ \left. \overline{\left\vert \mathcal{M}_{s_{2}s_{1}s_{1}^{\prime}s_{2}^{\prime}}(I_2,I_1,I_N,I_\Xi)\right\vert^{2}}\right]\nonumber
\end{eqnarray}

where $s=(p_1+p_2)^2$ and $t=(p_1-p_3)^2$ are the usual squared
center-of-mass energy of the colliding hyperons and the squared four
momentum transfer in the reaction; $p_i$ are the momenta of the
ingoing hyperons in their center of mass frame, $s_{1,2,1^\prime,2^\prime}$ and $I_{1,2,Y,N,\Xi}$ are the spins and isospins of the incoming hyperons and outgoing nucleon and $\Xi$.
The explicit, iso-spin dependent Matrix elements and scattering amplitudes are given in appendix \ref{ap1}.
Combining the coefficients listed in equations(\ref{iso_2}) with eq. (\ref{crs_2}), we then obtain the cross sections of all possible reaction channels. 
The resulting crossections for the iso-spin dependent channels then are parametrized as:

\begin{widetext}
\begin{eqnarray}
	\sigma_{\Lambda \Lambda \rightarrow \Xi^- p}=\sigma_{\Lambda \Lambda \rightarrow \Xi^0 n}&=&\frac{1}{2}\sigma_{\Lambda \Lambda \rightarrow \Xi N}=\frac{37.15}{2}  \frac{p_N}{p_{\Lambda}}(\sqrt{s}-\sqrt{s_0})^{-0.16} \rm{mb}\\ 
		\sigma_{\Lambda \Sigma^+ \rightarrow \Xi^0 p}=\sigma_{\Lambda \Sigma^- \rightarrow \Xi^- n}&=&24.3781 (\sqrt{s}-\sqrt{s_0})^{-0.479} \rm{mb}\\
		\sigma_{\Lambda \Sigma^0 \rightarrow \Xi^- p}=\sigma_{\Lambda \Sigma^0 \rightarrow \Xi^0 n}&=&
\begin{cases}
6.475   (\sqrt{s}-\sqrt{s_0})^{-0.4167} \rm{mb} & \rm{for } \ (\sqrt{s}-\sqrt{s_0})<0.03336 \rm{GeV} \\
14.5054 (\sqrt{s}-\sqrt{s_0})^{-0.1795} \rm{mb} & \rm{for } \ (\sqrt{s}-\sqrt{s_0})>0.03336 \rm{GeV} 
\end{cases}\\
		\sigma_{\Sigma^0 \Sigma^0 \rightarrow \Xi^- p}=\sigma_{\Lambda \Sigma^0 \rightarrow \Xi^0 n}&=&
\begin{cases}
5.625  (\sqrt{s}-\sqrt{s_0})^{-0.318} \rm{mb} & \rm{for } \ (\sqrt{s}-\sqrt{s_0})<0.09047 \rm{GeV} \\
4.174 (\sqrt{s}-\sqrt{s_0})^{-0.4421} \rm{mb} & \rm{for } \ (\sqrt{s}-\sqrt{s_0})>0.09047 \rm{GeV} 
\end{cases}\\
	\sigma_{\Sigma^+ \Sigma^0 \rightarrow \Xi^0 p}=\sigma_{\Sigma^0 \Sigma^- \rightarrow \Xi^- n}&=& 4 \sigma_{\Sigma^0 \Sigma^0 \rightarrow \Xi^- p}\\
	\sigma_{\Sigma^+ \Sigma^- \rightarrow \Xi^- p}=\sigma_{\Sigma^+ \Sigma^- \rightarrow \Xi^0 n}&=&14.194 (\sqrt{s}-\sqrt{s_0})^{-0.442} \rm{mb}
\end{eqnarray}

\end{widetext}

and they are also presented in figure \ref{f2} as function of the two particle (hyperon) center-of-mass energy. As one can see, there is a significant isospin dependence in the exchange reactions and all, except the $\Lambda+\Lambda$ reaction, are exothermal.

\section{Results}
\begin{figure}[t]	
\center \includegraphics[width=0.5\textwidth]{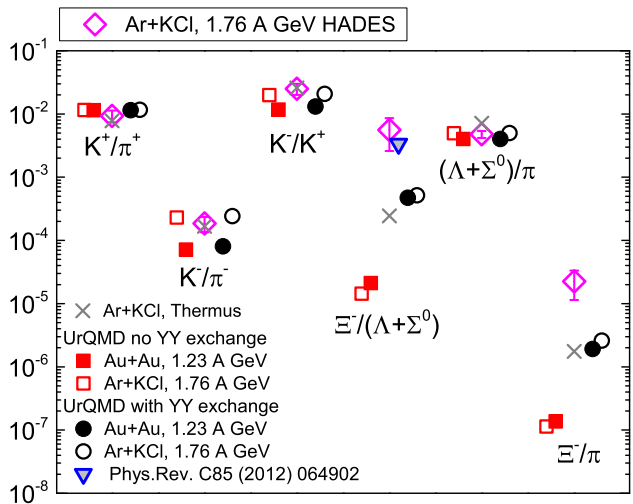}	
\caption{(Color Online) UrQMD results on strange particle ratios with (black squares) and without (red circles) the hyperon+hyperon strangeness exchange reaction.
we compare our results with HADES data \cite{Agakishiev:2009rr,Agakishiev:2010rs,Agakishiev:2009ar,Tlusty:2009dk}, a thermal fit to the HADES data on Ar+KCl at 1.76 A GeV (grey crosses, \cite{Agakishiev:2010rs,Wheaton:2004qb}) and a previous study employing the same crossections as in our study \cite{Li:2012bga}. }
\label{f3}
\end{figure}

After implementing the above cross sections, we apply the UrQMD model in its standard settings, i.e. only binary elastic and inelastic and $2\rightarrow n$ inelastic reactions (plus resonance decays) without nuclear potentials (no nuclear equation of state as e.g. in \cite{Li:2005kqa,Li:2005gfa,Li:2006ez}).
We investigate nuclear collisions of Ar+KCl at $E_{lab}= 1.76$ A GeV with an impact parameter of $b<5$ fm and Au+Au collisions at $E_{lab}= 1.23$ A GeV with an impact parameter of $b<9.5$ fm according to specifications from the HADES experiment. Comparisons of strange particle production with presently available HADES data are shown in figure \ref{f3}. Here we compare results where we only allow for the $\overline{K}+N \leftrightarrow \pi +Y$ exchange reaction with those where we also allow the $Y+Y\leftrightarrow \Xi + N $ exchange reaction. It is clearly visible, that the $\overline{K}+N\leftrightarrow \pi +Y$ nicely describes the ratios of single strange particles, measured by the HADES collaboration \cite{Agakishiev:2009rr,Agakishiev:2010rs,Agakishiev:2009ar,Tlusty:2009dk}.

When including the hyperon+hyperon strangeness exchange, we find that the exchange reaction is  the dominant source of $\Xi^-$ at the investigated energies.  When compared to a thermal fit \cite{Agakishiev:2010rs,Wheaton:2004qb} (grey crosses), we also get a reasonable agreement between the transport simulation and all the thermal yields, indicating that the microscopic simulation yields particle numbers close to chemical equilibrium. The conclusion, that the $\Xi$ yield is close to its equilibrium value, is also supported by the fact that an increase of the $Y+Y \leftrightarrow \Xi + N$ crossections by a factor of 2 only leads to a mild increase of the $\Xi^-/\Lambda$ ratio of $10\%$.
When we compare our results for the Ar+KCl reaction (open symbols) with the Au+Au collisions (full symbols) at the lower beam energy of $E_{lab}= 1.23$ A GeV, a particularly interesting result of our study is, that apparently the $\Xi^-/\Lambda$ ratio does not change with beam energy, while the $K^-/\pi$ ratio does show a clear beam energy dependence.
Furthermore, we observe a large discrepancy between our result and the $\Xi^-$ data (roughly a factor of 10 in the $\Xi^-/\Lambda$) as well as to the earlier study by Li et. al. \cite{Li:2012bga}, where the Relativistic Vlasov-Uheling-Uhlenbeck (RVUU) transport model \cite{Ko:1987gp,Chen:2003nm} was used to calculate the $\Xi/\Lambda$ ratio.

\section{Discussion}

\begin{figure}[t]	
\center \includegraphics[width=0.5\textwidth]{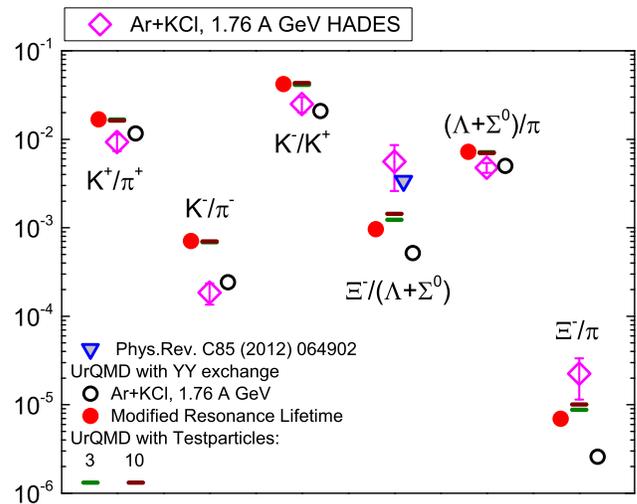}	
\caption{(Color Online) Comparing effects of different treatments of strangeness production in the transport model. Black open symbols are the results with standard parameters including the hyperon-hyperon strangeness exchange. The red filled symbols denote results when the resonance lifetime is set to essentially zero. The horizontal bars indicate the resulting ratios for the calculations with $n=3$ and $10$ test particles. Again HADES data on Ar+KCl at 1.76 A GeV are indicated as magenta diamonds.}
\label{f4}
\end{figure}

In the following section we will discuss the discrepancy between our results and the data and/or the earlier RVUU transport study.
One difference in the reaction dynamics, and specifically in the production of hyperons, between the two transport models (RVUU and UrQMD) is that RVUU allows for a direct associated production of $\Lambda$'s and kaons while in UrQMD, at this low energy, all $\Lambda$'s are produced via resonance decays. This leads to a delayed production of $\Lambda$'s and consequently a dilution of $\Lambda$ phase space density. To estimate the effect of a direct production mechanism we set the lifetime of baryonic non-strange resonances in UrQMD to essentially zero and recalculated all ratios. The results are presented in figure \ref{f4}. On can clearly see that the $\Xi/\Lambda$ ratio is increased significantly (as are many other ratios).\\
Another difference in the way the Boltzmann equation is solved in the two models is that UrQMD uses a microscopic geometrical interpretation of the scattering crossections with physical particles, while RVUU is based on the propagation and scattering of test particles. To estimate the effects of the systematic uncertainty, related to these different methods, we apply UrQMD in a test particle mode. This mode essentially multiplies the number of particles present in the collision by $n$, while all the scattering cross sections are consequently divided by $n$. The results for the particle ratios with 3 and 10 test particles are also shown in figure \ref{f4} as horizontal bars. Again the test particle method increases the $\Xi/\Lambda$ and $\Xi/\pi$ ratios, but leaves all other ratios constant, indicating that these ratios, not including the $\Xi$, are already close to their equilibrium values. Note that an effect of the test particle method on $\Xi$ production is, that the rarely produced hyperons can now rescatter with another hyperon test particle, even in collisions where only a single real hyperon is produced (self interaction).\\
Another difference between our approach and the one employed by Li et. al. \cite{Li:2012bga} is that we did not use iso-spin averaged cross sections. At the moment we cannot exclude that the production yields of the different isospin states of the hyperons can have an influence on the exchange probabilities (i.e. when only two $\Sigma^-$ are produced, no $\Xi$ can be formed by a exchange process, while this would be possible in the case of isospin independent cross sections). 

Both effects together (the instant creation and test particle method) can therefore account for a large portion of the difference in the model calculations. However, this does still not explain the deviation from the experimentally measured yield.\\
Nevertheless, the systematic uncertainties in the $\Lambda$ production (instant or delayed) and the scattering processes appear to be significant. 
Note that we additionally checked the effect of the nuclear equation of state, used in the simulation, on the two hyperon strangeness exchange process. We find, in accordance with \cite{Li:2012bga}, that there is no significant dependency of the final $\Xi/\Lambda$ ratio on the equation of state used.

\section{Conclusion}

We implemented the Y-Y strangeness exchange reaction in UrQMD and compared with HADES data. The exchange process is the dominant channel for $\Xi$ production. The $\Xi^-/\Lambda$ ratio is still significantly smaller than what is observed with HADES. However, we observed that the ratio appears to be not energy dependent at the two sub-threshold energies investígated. Systematic uncertainties in the $\Lambda$ production and treatment of scattering processes can account for some but not the whole difference between the RVUU and UrQMD model predictions.\\
In conclusion we can state that the large $\Xi$ yield at HADES is not fully explained within our model which gives a similar (but slightly larger) $\Xi/\Lambda$ ratio as the thermal fit. We also observe that the before mentioned ratio is almost constant within the beam energy range of the HADES experiment. We also observed that the explicit treatment of strangeness production (direct or through resonance) has a significant effect on multi-strange hyperon production. 

\section{Acknowledgments}
We would like to thank Che-Ming Ko for helpful comments, Tetyana Galatyuk and Manuel Lorenz for discussions on the HADES data.
This work was supported by GSI and the Hessian initiative for excellence (LOEWE) through the Helmholtz International Center for FAIR (HIC for FAIR). The computational resources were provided by the LOEWE Frankfurt Center for Scientific Computing (LOEWE-CSC).

\appendix
\section{The Y+Y scattering Amplitudes}\label{ap1}
 The explicit scattering amplitudes ($A^m$) follow from:  
\begin{eqnarray}\label{amplitude}
 \mathcal{M}_{s_{1}s_{2}s_{1}^{\prime}s_{2}^{\prime}}(I_Y,I_N,I_\Xi)&=&A^t(I_{Y,N,\Xi})\mathcal M^t_{s_{1}s_{2}s_{1}^{\prime}s_{2}^{\prime}}(s,t)\nonumber \\ &+&A^u(I_{Y,N,\Xi})\mathcal M^u_{s_{1}s_{2}s_{1}^{\prime}s_{2}^{\prime}}(s,u),\nonumber \\ 
 \mathcal{M}_{s_{1}s_{2}s_{1}^{\prime}s_{2}^{\prime}}(I_1,I_2,I_N,I_\Xi)&=&A^t(I_{1,2,N,\Xi})\mathcal M^t_{s_{1}s_{2}s_{1}^{\prime}s_{2}^{\prime}}(s,t)\nonumber \\ &+&A^u(I_{1,2,N,\Xi})\mathcal M^u_{s_{1}s_{2}s_{1}^{\prime}s_{2}^{\prime}}(s,u),\nonumber
\end{eqnarray}
with $u=(p_1-p_4)^2$. Both  $ \mathcal M^t_{s_{1}s_{2}s_{1}^{\prime}s_{2}^{\prime}}(s,t)$ and $\mathcal M^u_{s_{1}s_{2}s_{1}^{\prime}s_{2}^{\prime}}(s,u)$ have been calculated and are shown in ref. \cite{Li:2012bga}. Therefore we obtain,
\begin{eqnarray}\label{crs_2}
\sigma_{Y Y\to N\Xi}&=&\frac{1}{64\pi sp_i^2}\frac 1 {(2s_1+1)(2s_2+1)}\nonumber\\
&\times&\sum_{s_{1}s_{2}s_{1}^{\prime }s_{2}^{\prime }}\int dt\left[ \eta ^{tt}|\mathcal{M}_{s_{1}s_{2}s_{1}^{\prime }s_{2}^{\prime }}^{t}|^2\right. \nonumber\\
&-&\eta ^{tu}\mathcal{M}_{s_{1}s_{2}s_{1}^{\prime }s_{2}^{\prime }}^{t}\mathcal{M}_{s_{1}s_{2}s_{1}^{\prime }s_{2}^{\prime }}^{u\ast }\nonumber\\
&-&\eta^{ut}\mathcal{M}_{s_{1}s_{2}s_{1}^{\prime }s_{2}^{\prime }}^{u}\mathcal{M}_{s_{1}s_{2}s_{1}^{\prime }s_{2}^{\prime }}^{t\ast }\nonumber \\
&+&\left.\eta ^{uu}|\mathcal{M}_{s_{1}s_{2}s_{1}^{\prime }s_{2}^{\prime }}^{u}|^2\right],
\end{eqnarray}
where the coefficients $\eta^{lm}$ can be calculated for $\Lambda Y\to N\Xi$, 
\begin{eqnarray}
 \eta^{lm}&=&   A^l(I_Y,I_N,I_\Xi)A^{m\ast}(I_Y,I_N,I_\Xi),
\end{eqnarray}
and for $\Sigma\Sigma\to N\Xi$,
\begin{eqnarray}
 \eta^{lm}&=&  \frac 1 2 A^l(I_1,I_2,I_N,I_\Xi)A^{m\ast}(I_1,I_2,I_N,I_\Xi)\nonumber \\
           &+&\frac 1 2 A^l(I_2,I_1,I_N,I_\Xi)A^{m\ast}(I_2,I_1,I_N,I_\Xi),\nonumber
           \end{eqnarray}
The isospin factor for each reaction can be read from the interaction terms of the Lagrangian:
\begin{eqnarray}\label{iso}
A^t_{\Lambda\Lambda\to N\Xi}(I_N,I_\Xi)&=&u_{I_N}\cdot\sigma_1\cdot\left(u_{I_\Xi}\cdot\sigma_3\right)^T,\nonumber\\
A^u_{\Lambda\Lambda\to N\Xi}(I_N,I_\Xi)&=&u_{I_\Xi}\cdot\sigma_3\cdot\sigma_1\cdot u_{I_N}^T,\nonumber\\
A^t_{\Lambda\Sigma\to N\Xi}(I_\Sigma,I_N,I_\Xi)&=&u_{I_N}\cdot\sigma_1\cdot\left(u_{I_\Xi}\cdot\sigma_3\cdot\tau_{I_\Sigma}\right)^T\nonumber\\
A^u_{\Lambda\Sigma\to N\Xi}(I_\Sigma,I_N,I_\Xi)&=&u_{I_\Xi}\cdot\sigma_3\cdot\sigma_1\cdot\left(u_{I_N}\cdot\tau_{I_\Sigma}\right)^T\nonumber\\
A^t_{\Sigma\Sigma\to N\Xi}(I_1,I_2,I_N,I_\Xi)&=&\left(u_{I_{N}}\cdot \tau _{I_{1}}\right) \cdot \sigma_{1}\cdot \left( u_{I_{\Xi}}\cdot \sigma _{3}\cdot \tau _{I_{2}}\right) ^{T}\nonumber\\
A^u_{\Sigma\Sigma\to N\Xi}(I_1,I_2,I_N,I_\Xi)&=& \left(u_{I_{\Xi }}\cdot \sigma _{3}\cdot \tau _{I_{1}}\right) \cdot \sigma_{1}\cdot \left( u_{I_{N}}\cdot \tau _{I_{2}}\right) ^{T}\nonumber
\end{eqnarray}
where $\sigma_i$ are the Pauli matrices, $u_+=(1,0)$, $u_-=(0,1)$, $\tau_+=(\sigma_1+i\sigma_2)/\sqrt{2}$, $\tau_-=(\sigma_1-i\sigma_2)/\sqrt{2}$, and $\tau_0=\sigma_3$.\\
As a result, we obtain the following isospin dependent coefficients:
\begin{eqnarray}\label{iso_2}
1&=&\eta^{tt}_{\Lambda \Lambda \rightarrow \Xi^- p}=\eta^{tt}_{\Lambda \Lambda \rightarrow \Xi^0 n}=\eta^{ut}_{\Lambda \Lambda \rightarrow \Xi^- p}=\eta^{ut}_{\Lambda \Lambda \rightarrow \Xi^0 n}\nonumber\\
 &=&\eta^{tu}_{\Lambda \Lambda \rightarrow \Xi^- p}=\eta^{tu}_{\Lambda \Lambda \rightarrow \Xi^0 n}=\eta^{uu}_{\Lambda \Lambda \rightarrow \Xi^- p}=\eta^{uu}_{\Lambda \Lambda \rightarrow \Xi^0 n},\nonumber\\
2&=&\eta^{tt}_{\Lambda \Sigma^{+} \rightarrow \Xi^0 p}=\eta^{tt}_{\Lambda \Sigma^{-} \rightarrow \Xi^- n}=\eta^{ut}_{\Lambda \Sigma^{+} \rightarrow \Xi^0 p}\nonumber \\ &=&\eta^{ut}_{\Lambda \Sigma^{-} \rightarrow \Xi^- n}=\eta^{tu}_{\Lambda \Sigma^{+} \rightarrow \Xi^0 p}=\eta^{tu}_{\Lambda \Sigma^{-} \rightarrow \Xi^- n}\nonumber \\ &=&\eta^{uu}_{\Lambda \Sigma^{+} \rightarrow \Xi^0 p}=\eta^{uu}_{\Lambda \Sigma^{-} \rightarrow \Xi^- n},\nonumber\\
1&=&\eta^{tt}_{\Lambda \Sigma^{0} \rightarrow \Xi^- p}=\eta^{tt}_{\Lambda \Sigma^{0} \rightarrow \Xi^0 n}=\eta^{uu}_{\Lambda \Sigma^{0} \rightarrow \Xi^- p}\nonumber \\ &=&\eta^{uu}_{\Lambda \Sigma^{0} \rightarrow \Xi^0 n},\nonumber\\
-1&=&\eta^{ut}_{\Lambda \Sigma^{0} \rightarrow \Xi^- p}=\eta^{ut}_{\Lambda \Sigma^{0} \rightarrow \Xi^0 n}=\eta^{tu}_{\Lambda \Sigma^{0} \rightarrow \Xi^- p}\nonumber \\ &=&\eta^{tu}_{\Lambda \Sigma^{0} \rightarrow \Xi^0 n},\nonumber\\
4&=&\eta^{tt}_{\Sigma^0 \Sigma^{+} \rightarrow \Xi^0 p}=\eta^{tt}_{\Sigma^0 \Sigma^{-} \rightarrow \Xi^- n}=\eta^{ut}_{\Sigma^0 \Sigma^{+} \rightarrow \Xi^0 p}\nonumber \\ &=&\eta^{ut}_{\Sigma^0 \Sigma^{-} \rightarrow \Xi^- n}=\eta^{tu}_{\Sigma^0 \Sigma^{+} \rightarrow \Xi^0 p}=\eta^{tu}_{\Sigma^0 \Sigma^{-} \rightarrow \Xi^- n} \nonumber \\ &=&\eta^{uu}_{\Sigma^0 \Sigma^{+} \rightarrow \Xi^0 p}=\eta^{uu}_{\Sigma^0 \Sigma^{-} \rightarrow \Xi^- n},\nonumber\\
4&=&\eta^{tt}_{\Sigma^- \Sigma^{+} \rightarrow \Xi^- p}=\eta^{tt}_{\Sigma^- \Sigma^{+} \rightarrow \Xi^0 n}=\eta^{uu}_{\Sigma^- \Sigma^{+} \rightarrow \Xi^- p}\nonumber \\ &=&\eta^{uu}_{\Sigma^- \Sigma^{+} \rightarrow \Xi^0 n},\nonumber\\
0&=&\eta^{tu}_{\Sigma^- \Sigma^{+} \rightarrow \Xi^- p}=\eta^{tu}_{\Sigma^- \Sigma^{+} \rightarrow \Xi^0 n}=\eta^{ut}_{\Sigma^- \Sigma^{+} \rightarrow \Xi^- p}\nonumber \\ &=&\eta^{ut}_{\Sigma^- \Sigma^{+} \rightarrow \Xi^0 n},\nonumber\\
1&=&\eta^{tt}_{\Sigma^0 \Sigma^{0} \rightarrow \Xi^- p}=\eta^{tt}_{\Sigma^0 \Sigma^{0} \rightarrow \Xi^0 n}=\eta^{uu}_{\Sigma^0 \Sigma^{0} \rightarrow \Xi^- p}\nonumber \\ &=&\eta^{uu}_{\Sigma^0 \Sigma^{0} \rightarrow \Xi^0 n}=\eta^{tu}_{\Sigma^0 \Sigma^{0} \rightarrow \Xi^- p}=\eta^{tu}_{\Sigma^0 \Sigma^{0} \rightarrow \Xi^0 n}\nonumber \\ &=&\eta^{ut}_{\Sigma^0 \Sigma^{0} \rightarrow \Xi^- p}=\eta^{ut}_{\Sigma^0 \Sigma^{0} \rightarrow \Xi^0 n}.
\end{eqnarray}


			\end{document}